\documentclass[showpacs,prb,superscriptaddress,twocolumn]{revtex4-1}
\usepackage{graphicx}
\usepackage{enumerate}
\usepackage{hyperref}
\usepackage{textcomp}
\usepackage{amsmath}
\usepackage{amssymb}
\usepackage{pgf}
\usepackage{relsize}
\usepackage{pgfpages}
\usepackage{pgfplots}
\usepackage{tikz}
\usepackage{xifthen}
\usepackage{soul} 
\usetikzlibrary{quotes,angles}
\usetikzlibrary{arrows}
\usetikzlibrary{shapes.misc,shadows}
\usetikzlibrary{shapes,arrows}
\usetikzlibrary{calc,decorations.markings}
\usetikzlibrary{patterns}
\definecolor{myred}{rgb}{0.7,0.1,0.1}
\definecolor{myblue}{rgb}{0,0.447,0.741}
\definecolor{mygreen}{rgb}{0,0.498,0}
\definecolor{silvergray}{rgb}{0.752941176,0.752941176,0.752941176}
\definecolor{CateRed}{RGB}{161,24,1}

\graphicspath{{figures}}

\newcommand{\opcdag}[1]{{\hat{c}^{\dagger}}_{#1}}
\newcommand{\opc}[1]{{\hat{c}^{\phantom \dagger}}_{#1}}
\newcommand{\Spin}[2]{S_{#1}^{#2}}

\newcommand{\avg}{\mathrm{\scriptscriptstyle avg}}
\newcommand{\opt}{\mathrm{\scriptscriptstyle min}}
\newcommand{\PauliSigma}{\hat{\sigma}}
\newcommand{\Ham}{\widehat{H}}

\newcommand{\ud}{\mathrm{d}}
\newcommand{\nep}{\mathrm{e}}
\newcommand{\varepsres}{\varepsilon_\mathrm{res}}
\newcommand{\epsres}{\epsilon_\mathrm{res}}
\newcommand{\epsc}{\varepsilon_\mathrm{c}}
\newcommand{\epsgs}{\epsilon_\mathrm{gs}}
\newcommand{\cl}{\mathrm{cl}}
\newcommand{\Nsize}{L}
\newcommand{\tann}{\tau}

\begin{document}

\title{On the dynamics of Simulated Quantum Annealing in random Ising chains}

\author{Glen Bigan Mbeng}
\affiliation{SISSA, Via Bonomea 265, I-34136 Trieste, Italy}
\author{Lorenzo Privitera}
\affiliation{SISSA, Via Bonomea 265, I-34136 Trieste, Italy}
\affiliation{Institute of Theoretical Physics and Astrophysics, University of W\"urzburg, 97074 W\"urzburg, Germany
}
\author{Luca Arceci}
\affiliation{SISSA, Via Bonomea 265, I-34136 Trieste, Italy}
\author{Giuseppe E. Santoro} 
\affiliation{SISSA, Via Bonomea 265, I-34136 Trieste, Italy}
\affiliation{ICTP, Strada Costiera 11, 34151 Trieste, Italy}
\affiliation{CNR-IOM Democritos National Simulation Center, Via Bonomea 265, I-34136 Trieste, Italy}
%
%

\begin{abstract}
Simulated Quantum Annealing (SQA), that is emulating a Quantum Annealing (QA) dynamics on a classical computer 
by a Quantum Monte Carlo whose parameters are changed during the simulation, is a well established computational strategy 
to cope with the exponentially large Hilbert space.  
It has enjoyed some early successes but has also raised more recent criticisms.
Here we investigate, on the paradigmatic case of a one-dimensional transverse field Ising chain, two issues related to SQA in its
Path-Integral implementation: the question of Monte Carlo vs physical (Schr\"odinger) dynamics and the issue of the 
imaginary-time continuum limit to eliminate the Trotter error. 
We show that, while a proper time-continuum limit is able to restitute the correct Kibble-Zurek scaling of the residual energy
$\varepsres(\tau)\sim \tau^{-1/2}$ for the ordered case $---$ $\tau$ being the total annealing time ---, the presence of disorder leads to a 
characteristic {\em sampling crisis} for a large number of Trotter time-slices, in the low-temperature ordered phase. 
Such sampling problem, in turn, leads to SQA results which are apparently unrelated to the coherent Schr\"odinger QA even at intermediate $\tau$.   
\end{abstract}
\maketitle

\section{Introduction}
Quantum Annealing (QA) \cite{Finnila_CPL94,Kadowaki_PRE98,Brooke_SCI99,Santoro_SCI02,Santoro_JPA06}
--- essentially equivalent to a form of quantum computation known as Adiabatic Quantum Computation (AQC) \cite{Farhi_SCI01} ---
was originally introduced as an alternative to classical simulated annealing \cite{Kirkpatrick_SCI83} (SA) for optimization.
%
Due to the realisation of {\it ad-hoc} quantum hardware implementations, mainly based on superconducting flux qubits, QA is nowadays a 
field of quite intense research~\cite{Harris_PRB10, Johnson_Nat11, Denchev_2016, Lanting_2014, Boixo_2013, Dickson_2013, Boixo_2014}.
%
%
There are a number of important issues, both theoretical and experimental, related to QA, 
such as the question of a quantum speedup \cite{Ronnow2014_arxiv,Muthukrishnan2016,Albash_PRX18}, 
the role of ``non-stoquastic'' terms in the Hamiltonian \cite{Hormozi_PRB17,Susa2017}, 
or the effects due to the environment \cite{Amin_MRT_PRL08,Arceci_PRB17,Arceci_arxiv18}. 

At the theory level, the dynamics of a time-dependent quantum system under the action of a dissipative environment is a formidable problem. 
Even disregarding the effects of the environment, a detailed description of the unitary Schr\"odinger dynamics of a 
time-dependent quantum system --- for instance an Ising spin glass with classical Hamiltonian 
$\hat{H}_{\rm cl}(\PauliSigma^z_1 \cdots \PauliSigma^x_N)$ supplemented by a transverse field driving term:
\begin{equation} \label{eqn:Schroedinger}
i\hbar \frac{\ud}{\ud t} |\psi(t)\rangle = \bigg( \hat{H}_{\rm cl} - \Gamma(t) \sum_{i=1}^\Nsize \PauliSigma^x_i \bigg)  |\psi(t)\rangle \;,
\end{equation}
is usually limited to very small systems \cite{Kadowaki_PRE98,Farhi_SCI01}, not representative of the actual difficulty 
of a realistic problem.  
This has led, since the early days of QA \cite{Finnila_CPL94, Santoro_SCI02}, to QA-approaches employing imaginary-time 
Quantum Monte Carlo (QMC) techniques \cite{Becca_Sorella:book} 
--- most notably Path-Integral Monte Carlo (PIMC) \cite{Ceperley_RMP95} and Diffusion Monte Carlo \cite{Becca_Sorella:book} ---, 
at least in the most often considered ``stoquastic'' case, in which off-diagonal matrix elements of the Hamiltonian $\hat{H}(t)$ are 
non-positive. 
These approaches are generally known as Simulated QA (SQA), in analogy with classical Simulated Annealing (SA) \cite{Kirkpatrick_SCI83}.

In the case of a PIMC, SQA works as follows \cite{Santoro_SCI02,Martonak_PRB02}: 
one simulates the quantum system at a fixed value of the transverse field $\Gamma=\Gamma_0$ 
by resorting to a Suzuki-Trotter path-integral \cite{Suzuki_PTP76},  
which involves mapping the equilibrium quantum partition function, 
$Z_{Q}=\mathrm{Tr} \left[ \nep^{-\beta (\hat{H}_{\rm cl}-\Gamma\sum_i \PauliSigma^x_i)}\right]$, 
into the partition function of an equivalent classical Ising system with $P$ replicas of the original lattice. 
In principle one should take $P\to \infty$, a limit in which the mapping becomes exact.
Then, during the SQA simulation, the value of the transverse field $\Gamma(t)$ is decreased step-wise 
as a function of the {\em Monte Carlo time} $t$ down to a final (small) value. 

This approach raises a number of issues. 
On one hand, SQA is built on a classical Markov-chain dynamics 
which is in principle {\em unrelated} to the Schr\"odinger quantum dynamics of a real QA device; 
as usual, the Monte Carlo dynamics comes with a certain freedom in the choice of the Monte Carlo moves: what is the role of the moves chosen? 
On the other hand, the Suzuki-Trotter imaginary-time discretization would require taking the so-called {\em time-continuum limit} $P\to \infty$; 
however, if you think SQA as a classical optimization algorithm, then one might be interested in finding the optimal value 
\cite{Santoro_SCI02,Martonak_PRB02} of $P$, so to achieve the best  performance of the algorithm.
Quite evidently, the role of the $P\to \infty$ limit looses part of its meaning unless the SQA dynamics has something to do with the 
actual physical dynamics. 

Concerning the Monte Carlo vs physical dynamics issue, some initial evidence on ground state success probability histograms for Ising problems
\cite{Boixo_NatPhys13} encouraged to believe that SQA might have something to do with the actual QA dynamics of a real-world 
hardware: indeed, a certain degree of correlation between the performance of SQA and that of the D-Wave One QA device on random Ising 
instances with $\Nsize=108$ qubits was found.    
Equally encouraging was the message of Ref.~\onlinecite{Isakov_PRL16} (see also Ref.~\onlinecite{Mazzola_PRB17}) 
on the tunnelling rate between the two ground states of an ordered Ising ferromagnet: 
indeed, a correlation between the size-scaling of the PIMC tunnelling rate and the inverse squared gap $\Delta^{-2}$ 
calculated from exact diagonalization --- hence, likely, with the incoherent tunnelling rate of a real device --- was found.  
However, results which suggest a different scenario have meanwhile appeared in the literature: 
Ref.~\onlinecite{Amin_2017:arxiv} (see also Ref.~\onlinecite{Inack_PRA18}) studies the PIMC tunnelling rate in a frustrated toy model, 
showing that it does not match the inverse squared gap $\Delta^{-2}$. 
Refs.~\onlinecite{Albash_PRA15,Albash_EPJ15} show that the distributions of excited states and the qubit tunnelling spectroscopy data
observed in experiments with the D-Wave One QA device are not correctly reproduced by SQA. 
Finally, Ref.~\onlinecite{Albash_PRX18} demonstrates a scaling advantage of the last generation D-Wave chip against SA, while also showing that
a discrete-time ($P=64$) PIMC-SQA has an even better scaling.  

Concerning the time-continuum limit issue, Heim {\em et al.} \cite{Heim_SCI15} have pointed out that the optimization advantage of PIMC-SQA 
against classical SA, observed in Ref.~\onlinecite{Santoro_SCI02} for a suitably optimal finite value of $P$ in a two-dimensional random Ising model, 
might disappear when the limit $P\to \infty$ is properly taken.
But quite remarkably, as recently shown 
in Ref. \onlinecite{Zecchina_PNAS18}, non-convex optimization problems are known 
in which SQA, with the $P\to \infty$ limit properly taken, is definitely more efficient than its classical SA counterpart.  

Here we will reconsider these issues, trying to shed light onto some aspects of the fictitious Monte Carlo dynamics behind SQA.
We will do so by performing a detailed analysis of PIMC-SQA on a transverse-field random Ising spin chain, where exact equilibrium
and coherent-QA results are readily obtained by a Jordan-Wigner \cite{Lieb_AP61,Young1996} mapping to free spinless fermions. 
Due to the absence of frustration, we will compare PIMC-SQA results obtained with two types of Monte Carlo moves:  
Swendsen-Wang~\cite{Swendsen_PRL87} (SW) cluster moves limited to the imaginary-time direction, hence local in space, 
against space-time (non-local) SW cluster moves, which provides an extremely fast Monte Carlo dynamics.
We find that equilibrium thermodynamical PIMC simulations at finite $T$ clearly show a sampling problem emerging 
for large $P$ when local SW cluster moves limited to the time-direction 
--- the most natural candidate moves for a physical single-spin-flip dynamics --- are employed below the critical point 
$\Gamma < \Gamma_c$ and at low temperatures. 
Next, we move to comparing the annealing dynamics of SQA against coherent-QA evolution results performed by solving the time-dependent
Bogoljubov-de Gennes equations for the Jordan-Wigner fermions \cite{Caneva_PRB07}. 
We will show that, while the standard Kibble-Zurek $\tau^{-1/2}$ scaling~\cite{kibble80, zurek96, Polkovnikov_RMP11} of the residual energy 
is recovered in the ordered case, in presence of disorder the situation is more complex. 
The SQA dynamics shows a very interesting feature: the residual energy at $\Gamma(t)$ is essentially predicted by the corresponding
{\em equilibrium} thermodynamical value, but at an effective temperature $T_{\rm eff}(\tann)>T$. 
This aspect is shared by the coherent-QA evolutions, which can also be described by a similar {\em Ansatz}. 
However, the overall behaviours of $T_{\rm eff}(\tann)$ in the two cases, or equivalently that of $\varepsres(\tann)$ vs $\tau$, 
are definitely unrelated. 

The paper is organized as follows. Section \ref{model:sec} presents the model we study, the random Ising chain in a transverse field,  
and briefly describes the methods used: exact Jordan-Wigner mapping to free fermions and PIMC. 
Section \ref{results:sec} contains our results, both at equilibrium (Sec.~\ref{equilibrium:sec}) and for QA (Sec.~\ref{annealing:sec}).
Section \ref{conclusions:sec}, finally, contains our concluding remarks.

\section{Model and methods \label{model:sec}}
We consider a random Ising model in one dimension (1D) with open boundary conditions. 
The Hamiltonian in presence of a time-dependent transverse field $\Gamma(t)$ is 
\begin{equation}\label{eq:1d_TFIM}
  \Ham(t) = -\sum_{i=1}^{\Nsize-1} J_{i}\PauliSigma^{z}_{i} \PauliSigma^{z}_{i+1} - \Gamma(t) \sum_{i=1}^{\Nsize} \PauliSigma^{x}_{i} \;,
\end{equation}
where $\PauliSigma^{x,y,z}_i$ are Pauli matrices at site $i$, $J_i$ are random bond couplings 
and $\Nsize$ is the number of spins in the chain. 
We assume the bond couplings $J_i$ to be uniformly distributed independent positive random variables, $J_i \in (0,1]$.
For $\Gamma=0$, because of the simple geometry of the system, disorder causes no frustration, and the optimization task is trivial:
The two degenerate ``classical'' ground states of the system are simply the ferromagnetic states $|\uparrow \uparrow \cdots \uparrow \rangle$ and 
$|\downarrow \downarrow \cdots \downarrow \rangle$, with a minimum energy (per spin), given by 
$\epsgs(\Gamma=0)=-\frac{1}{\Nsize}\sum_{i=1}^{\Nsize-1} J_{i}$.
Nevertheless, disorder alone is sufficient to make the annealing {\em dynamics} --- both classical \cite{Suzuki_JSTAT09,Zanca_PRB16} and 
quantum \cite{Dziarmaga_PRB06,Caneva_PRB07,Zanca_PRB16} --- rather complex. 
 
Path-Integral Monte Carlo (PIMC) is a standard approach to simulate the {\em equilibrium} properties of the Hamiltonian~\eqref{eq:1d_TFIM} 
at finite temperature $T>0$ when $\Gamma$ does not depend on time. It works as follows: 
We first apply a standard Suzuki-Trotter \cite{Suzuki_PTP76} mapping of the quantum system at a fixed temperature $T$,
corresponding to $\beta=1/(k_BT)$, into $P\to \infty$ classical coupled replicas:
\begin{equation} \label{eq:ZQ}
Z_{Q}=\mathrm{Tr} \, \nep^{-\beta \Ham} \simeq \lim_{P\to \infty} \sum_{S}^{\mathrm{config}} \nep^{-K_{\cl}[S]} \;,
\end{equation}
%
which interact with a classical action
\begin{equation} \label{eq:Kcl}
    K_{\cl} = -\sum_{k=1}^{P}\sum_{i=1}^{\Nsize-1} \, \left( \beta_{P} J_{i}\, \Spin{i}{k}\, \Spin{i+1}{k} + J^{\perp} \Spin{i}{k}\, \Spin{i}{k+1} \right) \;,
\end{equation}
at an effective temperature $PT$, corresponding to $\beta_{P}\equiv \beta/P \equiv \Delta \tau$.
Here $\Spin{i}{k}=\pm 1$ with $k=1\cdots P$ is a classical Ising spin at site $i$ and ``imaginary-time slice'' $\tau_k=(k-1)\beta/P=(k-1)\Delta \tau$,  
with boundary condition $\Spin{i}{P+1}\equiv \Spin{i}{1}$ required by the quantum trace in the partition function.
(The sum over configurations in Eq.~\eqref{eq:ZQ} runs over $S=\{S_i^k\}$.)   
The uniform ferromagnetic coupling $J^{\perp}$ along the imaginary-time direction is set by: 
\begin{equation}
    J^{\perp} =  -\frac{1}{2}\log \left[\tanh\left(\beta_{P} \Gamma \right)\right] \;.
\end{equation}
The correct quantum mechanical equilibrium calculation is recovered by taking the limit $P\to \infty$. 
Using a Metropolis algorithm we can then implement several different Monte Carlo dynamics for 
$K_{\cl}$, depending on the choice of the {\em Monte Carlo moves} on which the corresponding classical Markov chain is built. 
In an equilibrium PIMC, this would make no difference for the final equilibrium averages: it would just influence how fast the system reaches
the equilibrium steady state on which averages are calculated. 
In an annealing framework, the choice of the Monte Carlo moves is a delicate matter influencing the outcome of the SQA simulation.
Indeed, SQA is built by appropriately changing the transverse field $\Gamma$ during the course of the PIMC simulation in the hope of
mimicking the physical annealing dynamics behind Eq.~\eqref{eq:1d_TFIM}: there is no intrinsic separation between transient and stationary state. 
In the following, we will investigate and compare two different Monte Carlo moves:
\begin{description}
\item[1)] \textit{time cluster flips} (local in space). Given a site $i$, clusters of spins $\{\Spin{i}{k}\}$ are constructed along the imaginary-time direction
using the Swendsen-Wang algorithm \cite{Swendsen_PRL87}. This is the choice of Ref.~\onlinecite{Heim_SCI15}. 
A single Monte Carlo Step (MCS) consists of $\Nsize$ time-cluster flips.
\item[2)] \textit{space-time cluster flips} (non-local). Since the classical action in Eq.~\eqref{eq:Kcl} is ferromagnetic (unfrustrated), one can adopt
algorithms which construct space-time clusters, either Swendsen-Wang \cite{Swendsen_PRL87} or Wolff \cite{Wolff_PRL89}.   
In a single MCS one space-time cluster flip is performed.
\end{description}

The advantage of working with a random Ising chain is that {\em exact} equilibrium as well as coherent evolution QA results 
can be easily obtained and compared to PIMC data.
Indeed, using a Jordan-Wigner transformation, the Hamiltonian in Eq.~\eqref{eq:1d_TFIM} can be mapped to the following 
free-fermionic Hamiltonian
%
\begin{equation}
    \Ham = -\sum_{i=1}^{\Nsize-1}\, J_i (\opcdag{i} - \opc{i})(\opcdag{i+1} + \opc{i+1})
    - \Gamma\sum_{i=1}^{\Nsize} (2\opcdag{i} \opc{i}-1) \;,
\end{equation}
where $\opcdag{i}$ and $\opc{i}$ are spinless fermionic operators. 
In equilibrium --- $\Gamma$ independent of $t$ --- one can diagonalize such a BCS-like Hamiltonian by a Bogoliubov transformation, 
constructed by the numerical diagonalization of a $2\Nsize \times 2\Nsize$ matrix \cite{Young1996,Caneva_PRB07}.
The relevant quantity that we will consider is the difference (per spin) between the interaction energy's thermal average 
at a given value of $\Gamma$ and $T$ and the ferromagnetic classical ground-state energy $\epsgs(\Gamma=0)$:
\begin{equation} \label{eq:epsc}
    \epsc(\Gamma, T) =  \frac{1}{\Nsize} \sum_{i=1}^{\Nsize-1} J_{i} \,
    \Big( 1 - \langle \PauliSigma^{z}_{i} \PauliSigma^{z}_{i+1} \rangle_{\Gamma, T} \Big) \;.
\end{equation} 
$\epsc(\Gamma,T)$ quantifies thermal and quantum fluctuations over the classical ground states energy. 
Within a coherent-QA framework, where $\Gamma(t)$ is slowly switched to $0$ in a timescale $\tann$ and the Schr\"odinger dynamics
\eqref{eqn:Schroedinger} is followed, one can consider the time-dependent residual energy:
\begin{equation}\label{eq:interaction_energy per spin}
    \epsres(t, \tann) =  \frac{1}{\Nsize} \sum_{i=1}^{\Nsize-1} J_{i} \,
    \Big( 1 - \langle \PauliSigma^{z}_{i} \PauliSigma^{z}_{i+1} \rangle_{t} \Big) \;,
\end{equation} 
where now 
$\langle \PauliSigma^{z}_{i} \PauliSigma^{z}_{i+1} \rangle_{t} = \langle \psi(t) | \PauliSigma^{z}_{i} \PauliSigma^{z}_{i+1} |\psi(t) \rangle$
is the quantum average with the time-evolving state $|\psi(t)\rangle$. 
It can be calculated through time-dependent Bogoljoubov-de Gennes (BdG) equations \cite{Caneva_PRB07,Zanca_PRB16}. 
The residual energy at the end of the annealing is simply obtained as 
\begin{equation}
\varepsres(\tann) \equiv \epsres(t=\tann,\tann) \;.
\end{equation}

\section{Results} \label{results:sec}
We now discuss the results obtained on the random Ising chain problem. 
We start from the equilibrium thermodynamics at finite $T$ and $\Gamma$, where we compare the different choices of
Monte Carlo moves --- essentially, Swendsen-Wang cluster moves restricted to the time direction only, or extended to 
space and time --- on the way the exact results are attained for $P\to \infty$. 
Interestingly, we find that, in presence of disorder, there is a clear {\em sampling problem} for the time cluster moves as $P$ increases in the ferromagnetic phase $\Gamma < \Gamma_c$. 

Next, we move to comparing the annealing dynamics of SQA against coherent-QA evolution results performed by solving the time-dependent
Bogoljubov-de Gennes equations for the Jordan-Wigner fermions \cite{Caneva_PRB07}. 
We show that, while SQA recovers the standard Kibble-Zurek $\tau^{-1/2}$ scaling~\cite{kibble80,zurek96,Polkovnikov_RMP11} 
of the residual energy in the ordered case, in presence of disorder the situation is less clear. 

\subsection{Equilibrium PIMC simulations} \label{equilibrium:sec}
\begin{figure}
\includegraphics[width=8.0cm]{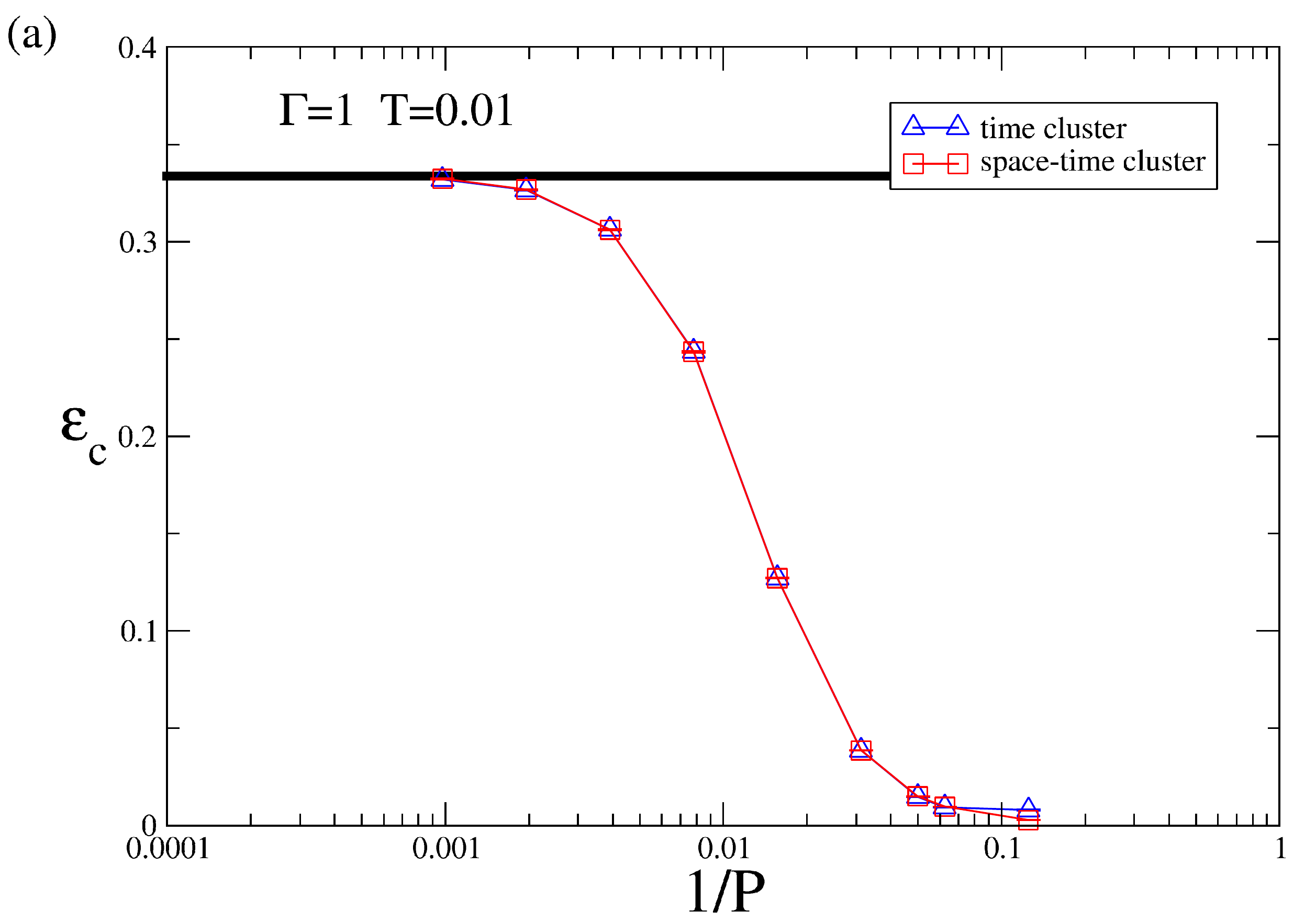}\\
\includegraphics[width=8.0cm]{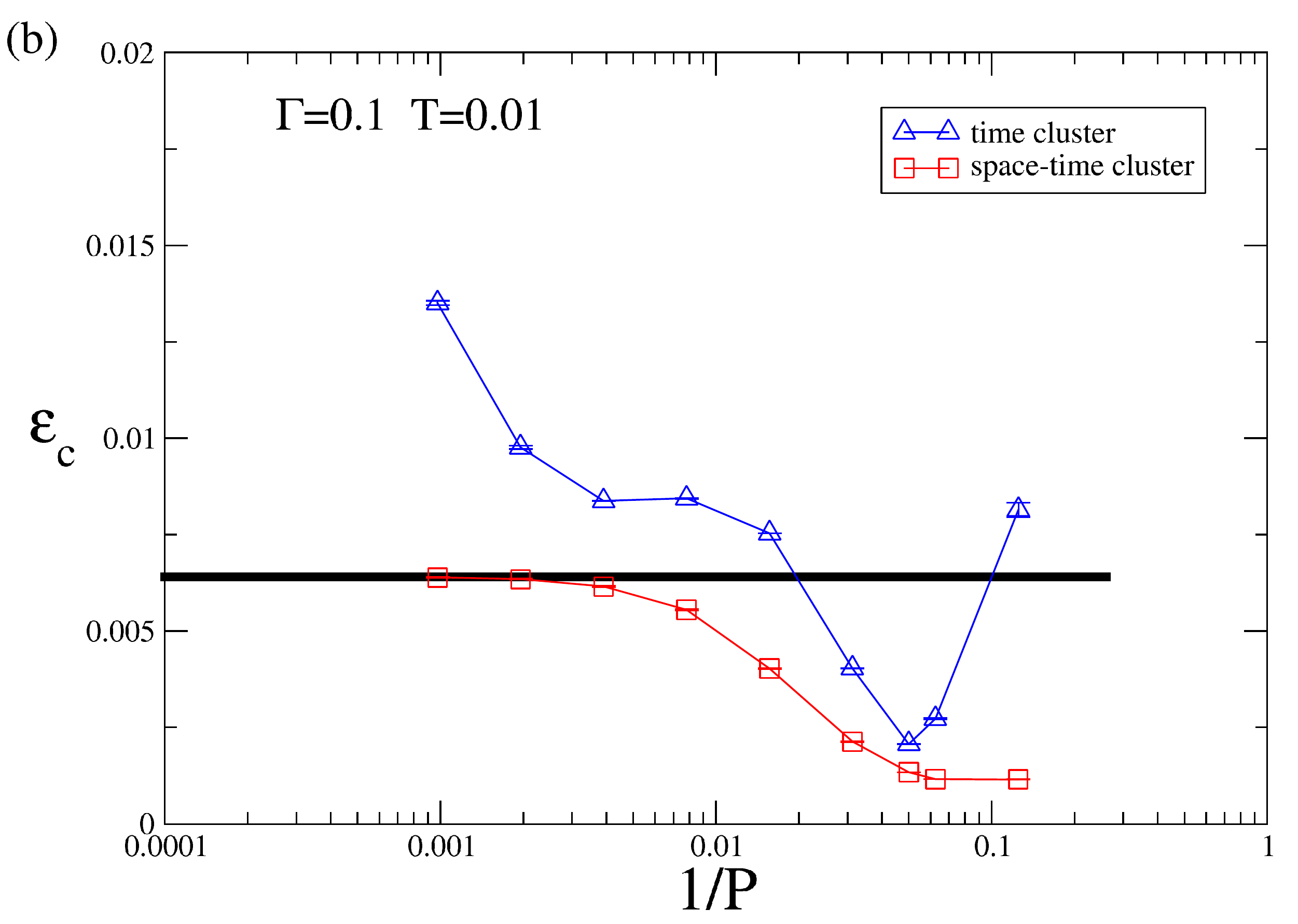}
\caption{PIMC estimates of $\epsc(\Gamma, T)$ in Eq.~\eqref{eq:epsc} at low temperature $k_BT=0.01$ and for two values 
of the transverse field, $\Gamma=1$ (a) and $\Gamma=0.1$ (b), as a function of the inverse number of 
Trotter slices $1/P$. The results are obtained for a given random instance of a chain of $\Nsize=256$ spins.
The horizontal thick lines denote the exact $\epsc(\Gamma, T)$ calculated from the Jordan-Wigner calculation. 
The simulation length is here $t_{\rm run}\lesssim 10^8\mbox{MCS}$. 
An initial $t_{\rm burn}$ MCS were discarded to ensure the equilibration of the Markov chain.
The value of $t_{\rm burn}$ was chosen using Geweke's diagnostic \cite{Brooks_1998}, while taking care to not discard
more than 50\% of the iterations.
}
\label{fig:eq_PIMC_gamma}
\end{figure}

Figure~\ref{fig:eq_PIMC_gamma} shows our PIMC equilibrium estimates for $\epsc(\Gamma,T)$ in Eq.~\eqref{eq:epsc}
at low temperature, $T=0.01$, for two values of $\Gamma$, above and below the $T=0$ quantum critical point \cite{Fisher_PRB95}, here
at $\Gamma_c=1/\nep$ with $J_i\in [0,1]$, which sets our energy scale.
The data shown refer to a {\em single realization} of disorder (no disorder average), in order to precisely test their convergence to the exact value for the same
realization of $\{J_i\}$: they are representative of all the instances we have tested.  
Results for both types of Monte Carlo moves are shown by triangles (time cluster moves) and squares (space-time cluster moves). 
For $\Gamma=1>\Gamma_c$, we see that both Monte Carlo moves provide consistent estimates of $\epsc(\Gamma, T)$, 
which approach from below the correct exact value, denoted by the horizontal line, as $P\to \infty$. 
Notice that, for finite $P$, the Trotter discretization error 
--- of order $\mathcal{O}(\frac{1}{P^2T^3})$, which amounts to a 10\% error at $P\approx 100$ for the case shown ---, 
introduces a bias towards lower values of $\epsc$. 
%
%
Even more interesting is the outcome for $\Gamma=0.1<\Gamma_c$, see Fig.~\ref{fig:eq_PIMC_gamma} (b). 
Here we see that the space-time (non-local) cluster moves correctly reproduce the exact $P\to \infty$ value, with the usual Trotter-error bias.
However, the time-cluster moves (local in space) {\em completely miss the exact target}: as $P$ increases, the PIMC value first seems to move
towards the exact one, up to roughly $P_*\sim 32 \div 64$, but than strongly overshoots the target and shows deviations as large as a 100\% error for the highest $P=1024$.
This implies that the time cluster moves are unable to correctly sample the correct distribution, especially at relatively large $P$, even with quite
long simulation times of order $t_{\rm run}\sim 10^8 \, \mbox{MCS}$. 
The fact that a large-$P$ Trotter sampling is definitely non-trivial is well known for PIMC in continuous systems, see for instance Ref.~\onlinecite{Brualla_JCP04}.
The time cluster moves are, however, the only candidate moves for PIMC in frustrated systems, where space-time cluster moves cannot be 
employed; they are also a quite natural implementation of a ``physical'' single-spin-flip dynamics. 
     
\subsection{PIMC-SQA compared to coherent QA} \label{annealing:sec}
\begin{figure}
   \includegraphics[width=8cm]{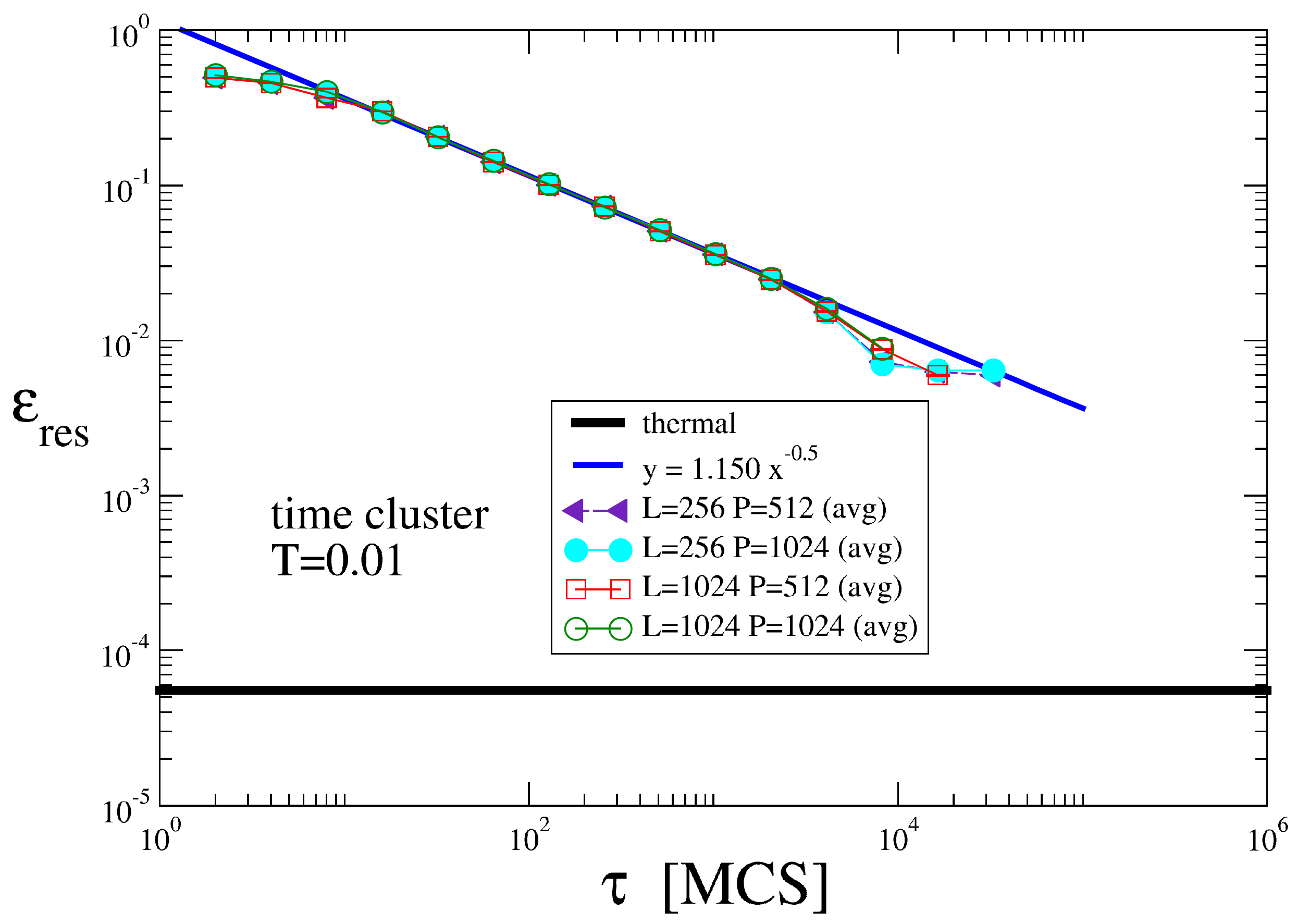}
   \caption{Test for Kibble-Zurek behaviour, $\varepsres(\tann) \sim \tann^{-1/2}$, in the ordered transverse-field Ising model. 
   SQA is here implemented with SW time cluster moves at $T=0.01$. 
   The horizontal thick line denotes the equilibrium thermal value of $\epsc(\Gamma=0,T=0.01)$.}
\label{fig:SQA_ordered}
\end{figure}

We now turn to the SQA dynamics. 
As done in previous studies \cite{Santoro_SCI02,Martonak_PRB02,Heim_SCI15}, we use a protocol in which $\Gamma$ is linearly 
reduced to zero as a function of the Monte Carlo time. 
More precisely, we start from $\Gamma(0)=2.5$ and perform a preliminary equilibration of the system.
We then reduce $\Gamma(t)$ at each MCS in such a way that $\Gamma(\tann)=0$, where $t$ is the time in MCS units 
and $\tann$ the total annealing time:
\begin{equation}
\Gamma(t)=\Gamma(0) \left(1-\frac{t}{\tann} \right) \;. 
\end{equation} 
Notice that, in our choice, we reduce $\Gamma$ at each MCS, by a rather small quantity $\Delta \Gamma=\Gamma(0)/\tau$,
rather than implementing a staircase with $N_{\Gamma}$ MCS at each of the $\tau/N_{\Gamma}$ values
of $\Gamma$: the results are essentially equivalent. 
Let us consider, as a warm up, the ordered case $J_i=J$, where we set $J=1$ to be our energy unit.
The coherent QA dynamics, here, is well known to produce a Kibble-Zurek~\cite{kibble80,zurek96,Polkovnikov_RMP11} 
decrease of the residual energy $\varepsres(\tann) \sim \tann^{-1/2}$.  
The SQA estimate, calculated from the Trotter replica average 
\begin{equation} \label{eq:eresavg}
    \varepsres^{\avg}(\tann) =  \frac{1}{\Nsize} \sum_{i=1}^{\Nsize-1} J_{i} \,
    \Big( 1 - \frac{1}{P} \sum_{k=1}^P \Spin{i}{k}  \Spin{i+1}{k}  \Big) \;,
\end{equation}
at the final configuration $\{ \Spin{i}{k} \}$, averaged over many repetitions of the SQA run, is shown in Fig.~\ref{fig:SQA_ordered} for
the time cluster moves.  
Quite remarkably, the behaviour of $\varepsres^{\avg}(\tann)$ is well compatible with the KZ coherent behaviour, with the
large $\tau$ deviations likely due to finite-size effects.  
We have verified that such KZ scaling would not be captured by using a SQA dynamics based on non-local (space-time) cluster moves.  
%

The natural question is whether this agreement survives also in the disordered case. 
\begin{figure}
   \includegraphics[width=8cm]{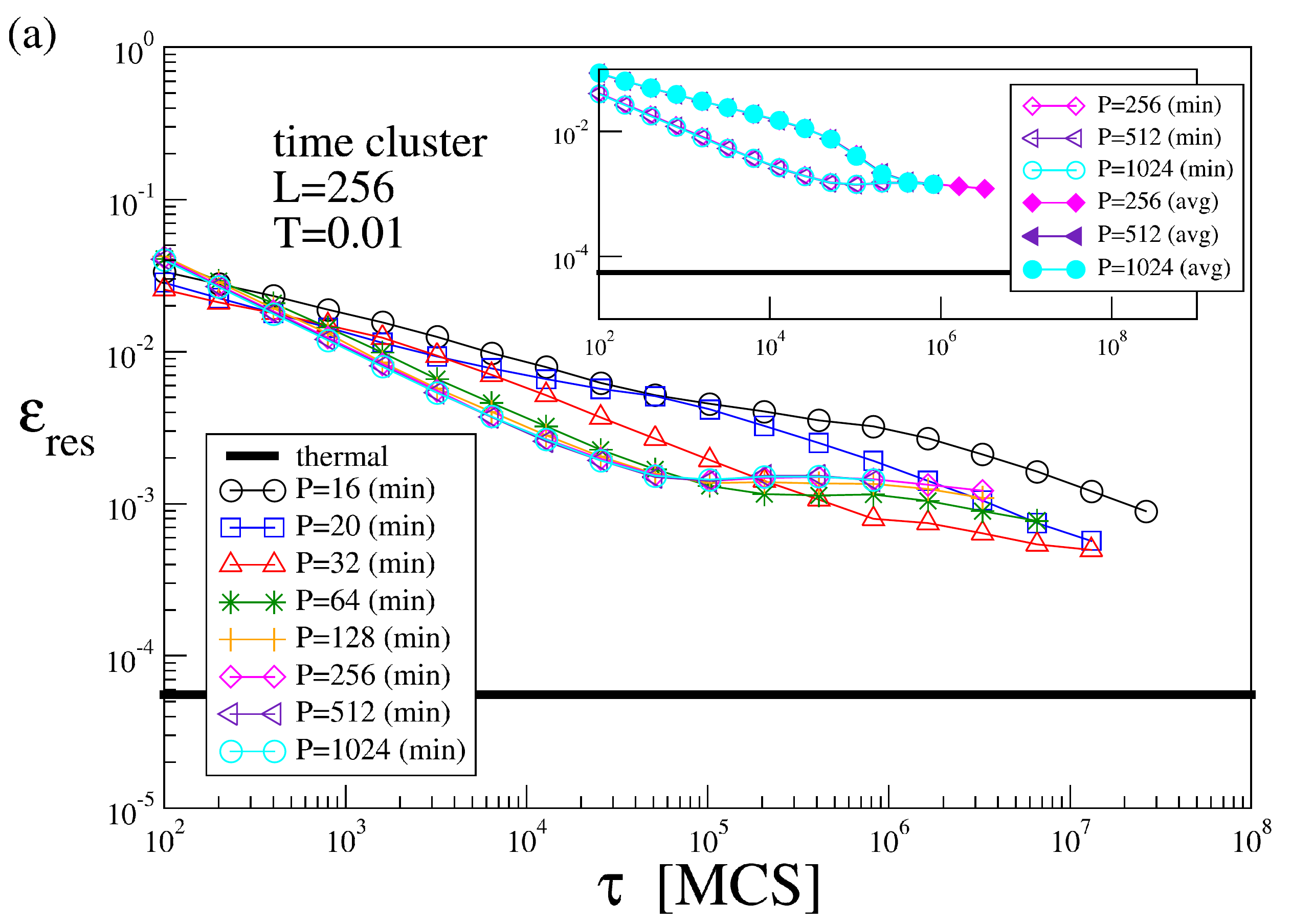}\\
   \includegraphics[width=8cm]{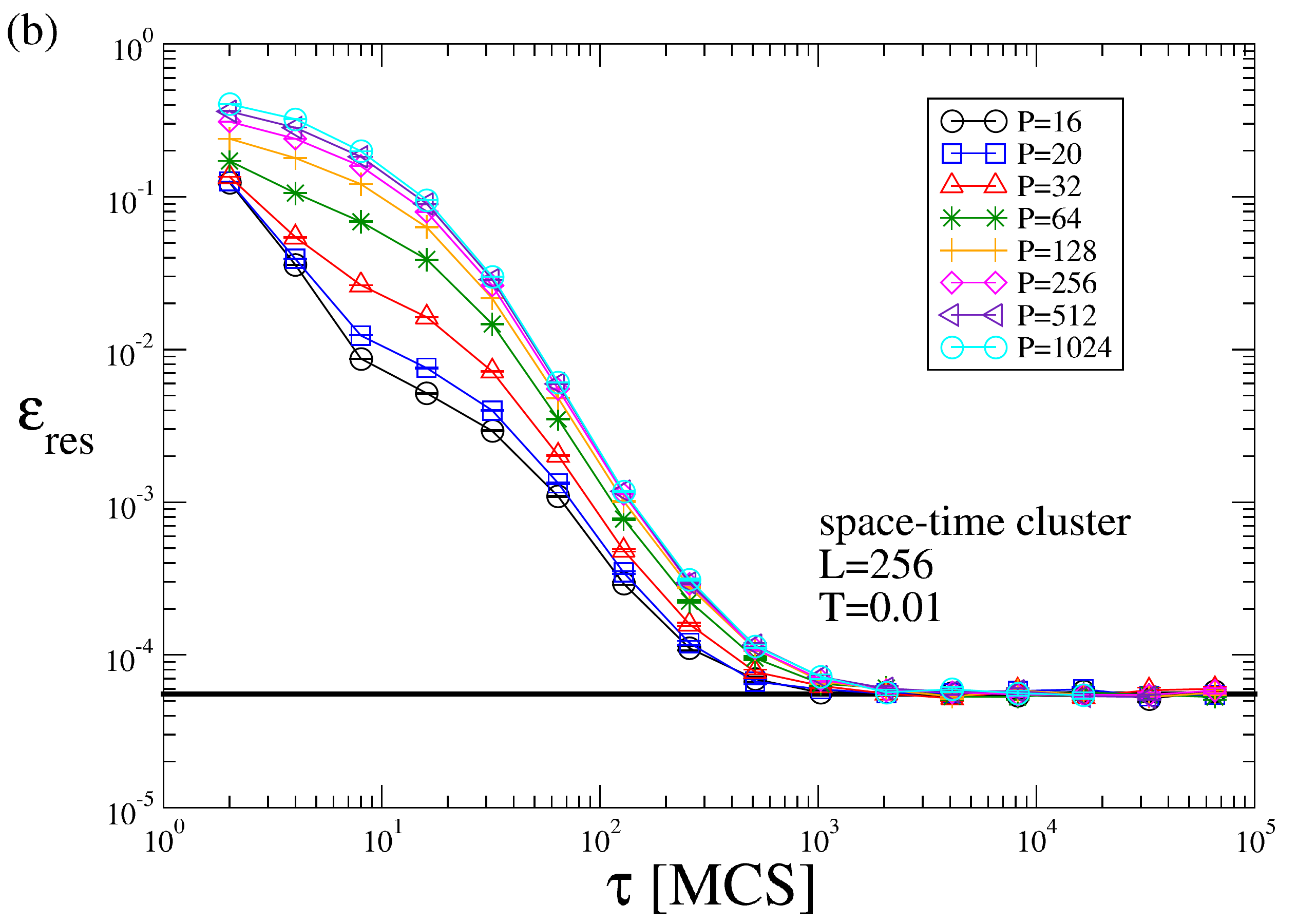}
   \caption{Residual energy at end of the SQA schedule as a function of the annealing time for various values of $P$ and a fixed disorder realization.
   The quantity we plot is $\varepsres^{\opt}(\tann)$, see Eq.~\eqref{eq:eresmin}, except in the inset, where $\varepsres^{\avg}(\tann)$ is also shown. 
   }
   \label{fig:SQA_T0.01}
\end{figure}
We start by showing the results obtained, in the same spirit of the SQA numerics presented 
in Refs.~\onlinecite{Santoro_SCI02,Martonak_PRB02,Heim_SCI15}, 
by considering the Trotter slice $k_{\star}$ that realizes the minimum classical energy value for the residual energy:
\begin{equation} \label{eq:eresmin}
    \varepsres^{\opt}(\tann) =  \frac{1}{\Nsize} \sum_{i=1}^{\Nsize-1} J_{i} \,  \Big( 1 - \Spin{i}{k_{\star}}  \Spin{i+1}{k_{\star}}  \Big) \;,
\end{equation}
for a given random instance of a chain with $\Nsize=256$ sites and $J_i\in [0,1]$. 
In Fig.~\ref{fig:SQA_T0.01}(a) we show SQA data obtained for various $P$ with the SW time cluster moves. 
Notice the strong similarity with the SQA data shown in Ref.~\onlinecite{Santoro_SCI02} and, in particular, with Fig.~3A of Ref.~\onlinecite{Heim_SCI15},
obtained for a two-dimensional frustrated Ising glass: this shows that, quite likely, the phenomena observed are due to disorder, rather than to a truly 
complex frustrated landscape. 
Notice also that, within an optimization framework, the optimal choice of $P$ is not $P\to \infty$, but rather $P_\opt\sim 32$, 
as indeed empirically found in Ref.~\onlinecite{Santoro_SCI02}. 
As pointed out in Ref.~\onlinecite{Heim_SCI15}, these results raise doubts whether any possible advantage of SQA over
plain SA might be lost in the proper quantum limit $P\to \infty$.  
Figure~\ref{fig:SQA_T0.01}(b) shows that the SQA results obtained with the SW space-time cluster moves behave in a completely
different way: they quickly converge to the expected thermal average $\epsc(\Gamma=0,T=0.01)$. 
This simply tells that the SQA results are highly sensitive to the type of MC moves one adopts, as perhaps expected: 
most likely, the space-time cluster non-local moves have little to do with any physical dynamics, as we will further comment on in the following. 

Returning to the time cluster SQA results, we re-plot them in the inset of Fig.~\ref{fig:SQA_T0.01}(a) for the largest $P$, to highlight the fact
that the $P\to\infty$ limit is indeed reached as soon as $P\ge 256$. 
Here, the two sets of data shown are $\varepsres^{\opt}(\tann)$, the optimal Trotter-slice value in Eq.~\eqref{eq:eresmin}, 
against the proper ``quantum average'' $\varepsres^{\avg}(\tann)$ in Eq.~\eqref{eq:eresavg}, which shows a much smoother 
and monotonic behaviour: notice that the two curves approach each other for the largest $\tann$ investigated. 
This witnesses the fact that, for these largest $\tann$, the quantum fluctuations 
--- {\em i.e.} the fluctuations along the Trotter-time direction --- 
seem to play no role towards the end of the annealing. 

\begin{figure}
        \includegraphics[width=8cm]{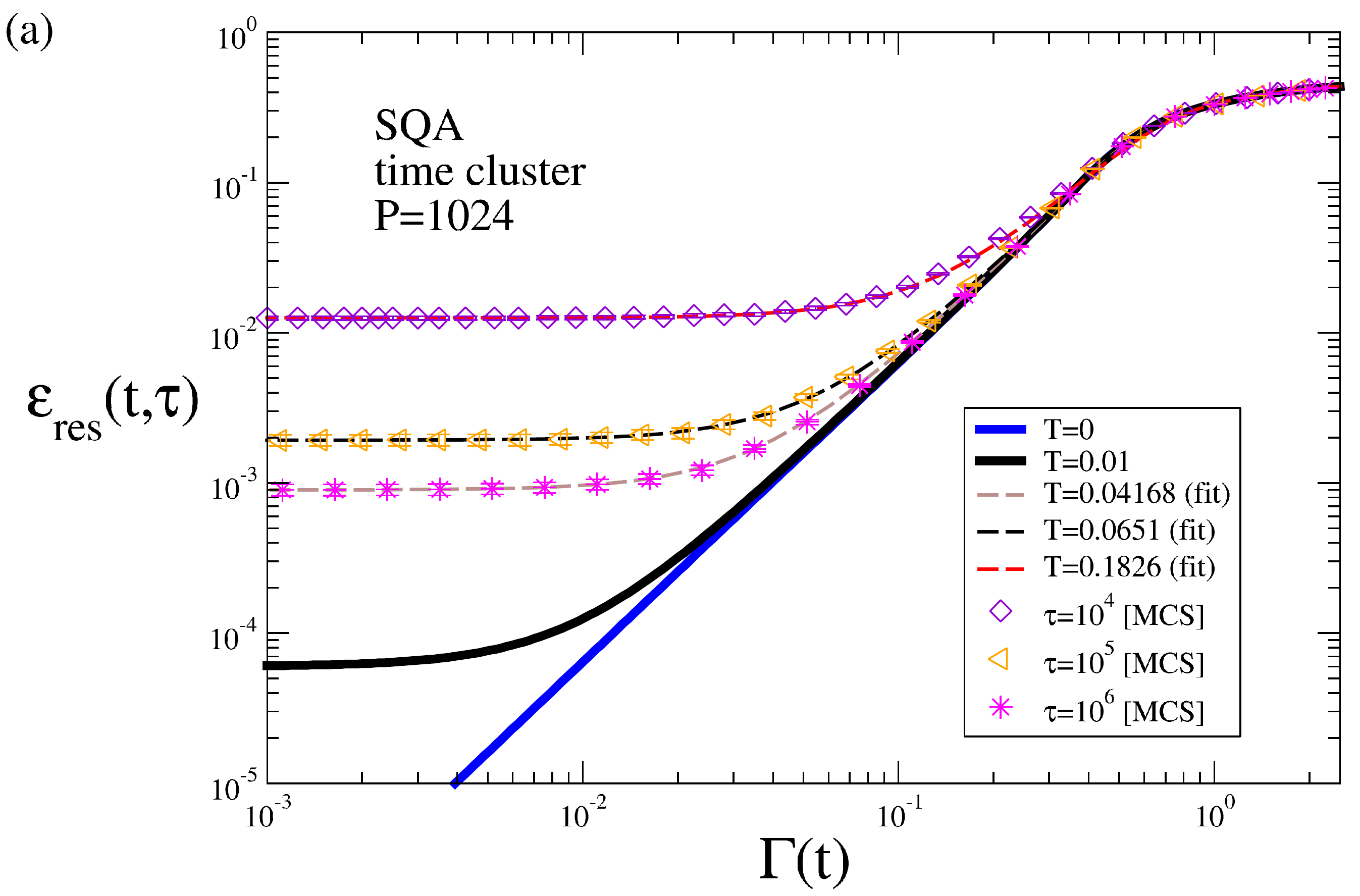}\\
        \includegraphics[width=8cm]{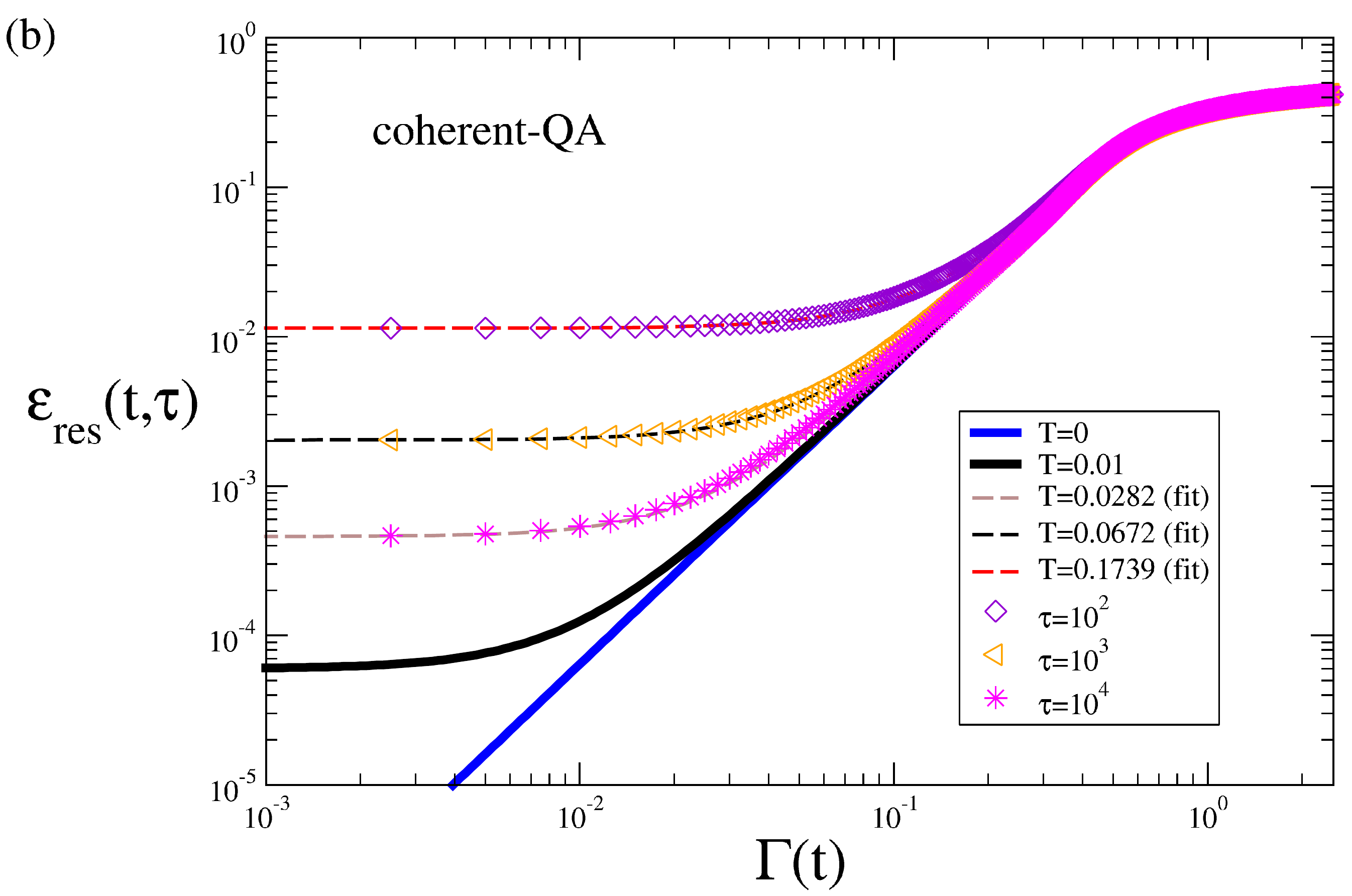}  
        \caption{
            Test of the dynamical {\em Ansatz} in Eq.~\eqref{eq:Ansatz} for the time-cluster SQA dynamics (a) and 
            the coherent QA dynamics (b), for the same disorder realization of Fig.~\ref{fig:SQA_T0.01}.
            The numerical data for $\epsres(t,\tann)$ are shown by points, for different $\tau$, at the corresponding value for $\Gamma(t)$, 
            while the fits with the equilibrium $\epsc(\Gamma,T_{\rm eff}(\tau))$ are shown by dashed lines. 
            For comparison, the exact equilibrium values for $\epsc(\Gamma,T=0.01)$ and $\epsc(\Gamma,T=0)$ are also shown by thick solid lines.}
            \label{fig:SQA_Ansatz}
\end{figure}

But the question remains: is there any physics that we can learn from the time cluster SQA dynamics in the disordered case? 
The first tests we have performed consist in monitoring the dynamics of $\epsres(t,\tau)$, for given $\tau$, versus $t$, 
both for the QA unitary evolution and the SQA dynamics. 
Indeed, since each $t$ is univocally associated to a value of $\Gamma(t)$, we can equivalently plot the SQA results, averaged over many repetitions
of the Monte Carlo dynamics, versus $\Gamma$. 
Figure~\ref{fig:SQA_Ansatz}(a) shows the results for three values of $\tau$, with the SQA results denoted by points. 
Here we find a surprising result: the SQA with time-cluster moves visits configurations which are essentially equilibrium configurations, but
at an effective temperature $T_{\rm eff}(\tann)$, which depends on the total annealing time $\tann$.
More precisely, we have verified that the following {\em Ansatz} for the dynamical residual energy holds:
\begin{equation} \label{eq:Ansatz}
\epsres(t,\tann) = \epsc(\Gamma(t),T_{\rm eff}(\tann)) \;,
\end{equation}
where the corresponding equilibrium values of $\epsc(\Gamma,T_{\rm eff})$, with $T_{\rm eff}$ obtained by fitting the numerical points, 
are shown by dashed lines in Fig.~\ref{fig:SQA_Ansatz}(a).
Even more remarkably, the same {\em Ansatz} also holds, on the same disordered instance, for the coherent QA dynamics,  
performed integrating through a $4^{\rm th}$-order Runge-Kutta algorithm the BdG equations~\cite{Caneva_PRB07,Zanca_PRB16} 
for the free-fermion Jordan-Wigner mapping, see Fig.~\ref{fig:SQA_Ansatz}(b).
We might go on and compare the corresponding $T_{\rm eff}(\tau)$ obtained for the two dynamics. 
However, since $\varepsres(\tau)=\epsc(\Gamma=0,T_{\rm eff}(\tann))$, due to the validity of the {\em Ansatz} \eqref{eq:Ansatz}, 
we can equivalently compare the results obtained for $\varepsres(\tau)$ in the two cases. 
\begin{figure}
       \includegraphics[width=8cm]{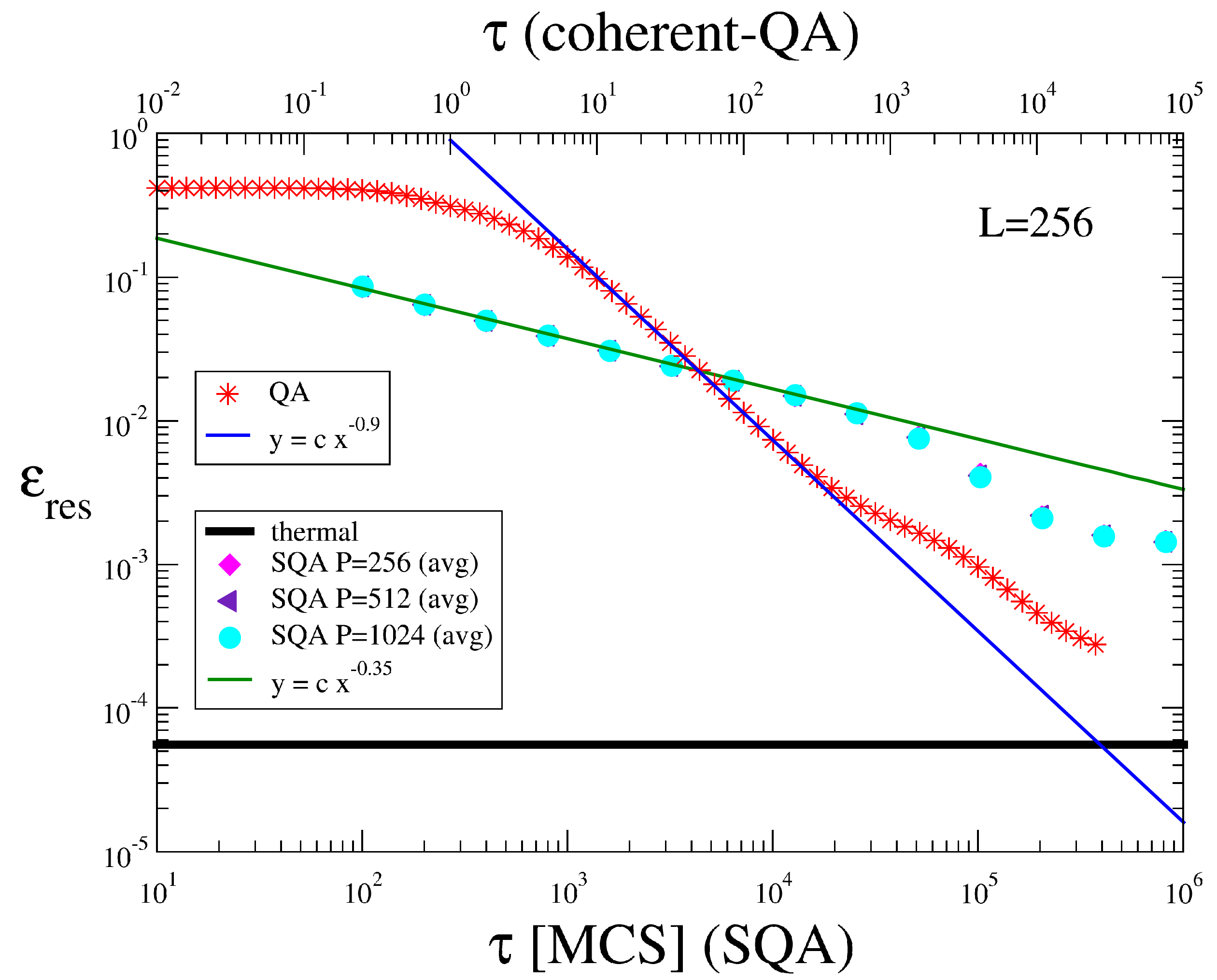}
       \caption{The residual energy $\varepsres(\tau)$ at the end of the annealing for the same disorder realization of Fig.~\ref{fig:SQA_T0.01}.
       The stars show the coherent-QA results vs $\tau$ (upper abscissa axis) while the other symbols refer to the SQA results (the same
       data shown in the inset of Fig.~\ref{fig:SQA_T0.01}(a)). 
       The equilibrium thermal value $\epsc(\Gamma=0,T=0.01)$ is shown by a thick horizontal line.
       }
       \label{fig:SQA_comparison}
\end{figure}
Figure \ref{fig:SQA_comparison} shows such a comparison. 
The solid symbols show the SQA results for $\varepsres^{\avg}(\tann)$ in Eq.~\eqref{eq:eresavg}, already reported in the inset of Fig.~\ref{fig:SQA_T0.01}(a), while the stars show the coherent-QA $\varepsres(\tann)$. 
Quite evidently, both plots show an intermediate power-law part which, however, shows markedly different power-law exponents 
in the two cases, $\sim \tau^{-0.9}$ for coherent-QA compared to $\sim \tau^{-0.35}$ for SQA: hence, no linear scaling of the physical 
against the MC time can ever make the two results consistent. 
The situation does not improve for large $\tau$, where it is known~\cite{Dziarmaga_PRB06,Caneva_PRB07} that, 
in the thermodynamic limit, the coherent-QA results would display a 
logarithmic slow-down~\cite{Santoro_SCI02}, $\varepsres(\tann)\sim [\log(\gamma \tau)]^{-\xi}$, with $\xi>2$: 
the data for $\Nsize=265$ evidently still suffer from 
finite-size effects which prevent from appreciating such a subtle logarithmic slow-down; nevertheless, they clearly depart from the 
intermediate-$\tau$ data by staying {\em above} the power-law $\sim \tau^{-0.9}$. 
On the contrary, the SQA data depart from their intermediate-time power-law $\sim \tau^{-0.35}$ from {\em below}, but then show a final 
slow-down of difficult interpretation. 

\section{Conclusions} \label{conclusions:sec}
We have investigated some aspects of the dynamics behind Simulated Quantum Annealing (SQA), specifically
its Path-Integral Monte Carlo (PIMC) implementation, through a detailed analysis of PIMC-SQA on a transverse-field random Ising spin 
chain, where exact equilibrium and coherent-QA results are easily obtained. 

Due to the absence of frustration, we were also able to compare results obtained with two types of Monte Carlo (MC) moves, a
local-in-space Swendsen-Wang cluster move limited to the imaginary-time direction, 
against space-time (non-local) SW cluster moves, which provides an extremely fast Monte Carlo dynamics.
The results show that the choice of the MC moves is of course crucial, but that fast non-local cluster moves have nothing to do
with any physical dynamics, which is better mimicked by local spin-flip moves.  

Concerning the latter more physical choice, we have verified that the expected Kibble-Zurek behaviour $\varepsres(\tann)\sim \tau^{-1/2}$ 
is well reproduced by SQA in the ordered case.
In presence of disorder, however, we found that equilibrium thermodynamical PIMC simulations at finite $T$ show a 
{\em sampling problem} emerging, for large $P$, below the critical point  $\Gamma < \Gamma_c$ and at low temperatures. 
The consequences of such a sampling problem on the SQA dynamics are {\em a priori} not obvious. 
Interestingly, we found that the time-dependent residual energy $\epsres(t,\tau)$ shows features that are shared also by
the coherent-QA Schr\"odinger dynamics, i.e., $\epsres(t,\tau)$ is perfectly described by the (instantaneous) equilibrium
value of $\epsc(\Gamma(t),T_{\rm eff}(\tau))$ at an effective temperature $T_{\rm eff}(\tau)$ which depends on the annealing time $\tau$.
Nevertheless, the SQA results for the residual energy $\varepsres(\tau)$ appear to be unrelated, in presence of disorder, with the corresponding 
coherent-QA results.

Several points still deserve a discussion. One might question the relevance of a comparison of SQA at a finite (low) $T$ against
QA results which assume a coherent Schr\"odinger evolution in absence of any external bath. On the practical side, we might add
that while a coherent-QA evolution is here quite easy to perform --- you just have to integrate $2\Nsize \times 2\Nsize$ BdG equations
for the Jordan-Wigner fermions --- the physical dissipative dynamics of a random Ising chain is still a problem which we do not know
how to efficiently and reliably tackle. More to the point, however, we can give arguments which are based on our current understanding
of the role of dissipation in the QA dynamics of the {\em ordered} Ising chain \cite{Patane_PRL08,Patane_PRB09,Arceci_arxiv18}. 
As indeed shown in Ref.~\onlinecite{Arceci_arxiv18}, and perhaps easy to argue about, dissipation has very little effect at small and intermediate
annealing times $\tau$, which implies that the different power-law behaviour displayed in Fig.~\ref{fig:SQA_comparison} 
would likely not be influenced by the presence of a bath. For large $\tau$, on the other hand, dissipation tends to drive the system closer
to a thermal steady state, which likely results in a larger residual energy, $\varepsres^{\rm diss}(\tau)>\varepsres(\tau)$, due to thermal 
defects generation.  
Hence, again, it is unlikely that the effect of a thermal bath at temperature $T$ would lead to a closer agreement between SQA and
a physical open quantum system dynamics. 

A few comments deserves also the largely accepted viewpoint that the time-continuum limit $P\to \infty$ is crucial for a comparison against
real QA hardware devices. While there is no question on the fact that, as pointed out in Ref.~\onlinecite{Heim_SCI15}, a
correct quantum mechanical treatment does require the limit $P\to \infty$, this by itself does not guarantee that the resulting SQA dynamics
is {\em physical}, as we have shown in this paper. 

The typical ``slow-down'' that SQA data with $P\to \infty$ tend to show for large $\tau$ should also not necessarily be taken to imply that 
there is no quantum speed-up of any type against classical Simulated Annealing (SA).
Indeed, based on theoretical arguments \cite{Santoro_SCI02}, a coherent-QA is expected to show some improvement in the exponent of the 
logarithmic scaling, $\varepsres(\tann)\sim [\log(\gamma \tau)]^{-\xi}$, against competing SA strategies: this improvement has been indeed 
verified on random Ising chains \cite{Caneva_PRB07,Suzuki_JSTAT09,Zanca_PRB16} and on infinitely connected $p$-spin Ising ferromagnets \cite{Wauters_PRA17}. 
Moreover, quite remarkably, non-convex optimization problems are known~\cite{Zecchina_PNAS18} in which SQA, 
with the $P\to \infty$ limit properly taken, is definitely much more efficient than its classical SA counterpart. 

\section*{ACKNOWLEDGMENTS}
We thank R. Fazio for discussions.
GES acknowledges support by the EU FP7 under ERC-MODPHYSFRICT, Grant Agreement No. 320796. 

\bibliography{BiblioSQA,BiblioQIC,BiblioQIsing2,BiblioQA,BiblioQIsing} 
\end{document}